\documentstyle[emulateapj]{article}

\newcommand{\kps}{\,\rm{km~s}^{-1}}

\newcommand{\muJy}{\,\mu\rm{Jy}}

\begin{document}

\lefthead{Frayer et al.}

\righthead{Identification of SMM\,J00266+1708}

\title{The Identification of the Submillimeter Galaxy SMM\,J00266+1708}

\author{D.\ T.\ Frayer\altaffilmark{1}, 
Ian Smail\altaffilmark{2},
R.\ J.\ Ivison\altaffilmark{3},
\&
N.\ Z.\ Scoville\altaffilmark{1}}

\altaffiltext{1}{Astronomy Department, California Institute of Technology
105--24, Pasadena, CA  91125, USA} 

\altaffiltext{2}{Department of Physics, University of Durham, South
Road, Durham, DH1 3LE, UK}

\altaffiltext{3}{Department of Physics \& Astronomy, University College
London, Gower Street, London, WC1E 6BT, UK}

\begin{abstract}

We report the detection of 1.3\,mm continuum and near-infrared $K-$band
($2.2\micron$) emission from the submillimeter galaxy SMM\,J00266+1708.
Although this galaxy is among the brightest sub-mm sources detected in
the blank-sky surveys ($L\sim 10^{13} L_{\sun}$), SMM\,J00266+1708 had
no reliable optical/near-infrared counter-part.  We used
sensitive interferometric 1.3\,mm observations with the Owens Valley
Millimeter Array to accurately determine the position of the sub-mm
galaxy.  Follow-up near-infrared imaging with the Keck~I telescope
uncovered a new faint red galaxy at $K=22.5$ mag which is spatially
coincident with the 1.3\,mm emission.  This is currently the faintest
confirmed counter-part of a sub-mm galaxy.  Although the redshift of
SMM\,J00266+1708 is still unknown, its high sub-mm/radio spectral index
suggests that the system is at high redshift ($z\ga2$).  Approximately
50\% or more of the sub-mm galaxies are faint/red galaxies similar to
that of SMM\,J00266+1708.  These ultraluminous obscured galaxies account
for a significant fraction of the total amount of star-formation at high
redshift despite being missed by optical/ultraviolet surveys.

\end{abstract}


\section{INTRODUCTION}

Deep surveys of the submillimeter sky using the Submillimeter Common
User Bolometer Array (SCUBA) camera (Holland et al. 1999) on the James
Clerk Maxwell Telescope have uncovered a population of ultraluminous
dusty galaxies at high-redshift (Smail, Ivison, \& Blain 1997; Hughes et
al.\ 1998; Barger et al.\ 1998; Eales et al.\ 1999; Blain et al.\
1999a).  This population accounts for a large fraction of the
extragalactic background at mm/sub-mm wavelengths (Blain et al.\ 1999b)
and hence is important to our understanding of the distant universe.
The sub-mm population is thought to contribute significantly to both the
total amount of star-formation (Blain et al.\ 1999b) and AGN activity
(Almaini et al.\ 1999) at high-redshift.  The sub-mm population will
likely show a mixture of AGN and starburst properties given their
apparent similarities to the local population of ultraluminous
($L>10^{12}L_{\sun}$) infrared galaxies (ULIGs, Sanders \& Mirabel
1996).  However, we could expect the majority ($\sim 70$--80\%) of the
sub-mm galaxies to be predominantly powered by starbursts since this has
been found for the local ULIGs (Genzel et al.\ 1998).  The early CO and
X-ray data on the sub-mm population support the starburst nature of the
population by showing the presence of sufficient molecular gas to fuel
the star-formation activity (Frayer et al.\ 1998, 1999) and the lack of
expected X-ray emission if mostly dominated by AGN (Fabian et al.\ 2000;
Hornschemeier et al.\ 2000).  Observations of the dust-rich sub-mm
galaxies complement the studies of the ultraviolet-bright Lyman-break
sources (Steidel et al.\ 1996, 1999) which tend to be much less luminous
at infrared wavelengths (Chapman et al.\ 2000).  Only by studying both
the Lyman-break and the sub-mm populations of galaxies will a complete
picture for the star-formation history of the universe emerge.

In order to understand the nature of the sub-mm population, we have been
carrying out multi-wavelength observations of individual systems in the
SCUBA Cluster Lens Survey (Smail et al.\ 1998).  This survey represents
sensitive sub-mm mapping of seven massive, lensing clusters which
uncovered 15 background sub-mm sources.  The advantage of this sample is
that the amplification of the background sources allows for deeper
source frame observations.  Also, lensing by cluster potentials does not
suffer from differential lensing so that the observed flux ratios will
represent intrinsic values, despite the possible variation of source
size at different wavelengths.

The most challenging aspect for follow-up observational studies of the
sub-mm population is determining the proper counter-parts to the sub-mm
emission and obtaining their redshifts (Ivison et al.\ 1998, 2000a).
The large 15$''$ SCUBA beam leaves ambiguity in identifying the galaxy
associated with the sub-mm emission.  The early results based on optical
imaging and spectroscopy were overly optimistic in the identification of
the sub-mm counter-parts (Smail et al.\ 1998; Barger et al.\ 1999; Lilly
et al.\ 1999).  Radio data (Smail et al.\ 2000a) and initial
near-infrared (NIR) imaging (Smail et al. 1999) suggest that several of
the original candidate optical counter-parts (e.g., Barger et al.\ 1999)
are incorrect.  Despite their ultra-high luminosities, many sub-mm
galaxies are nearly completely obscured by dust at ultraviolet/optical
wavelengths.  For these highly obscured galaxies, follow-up radio (Smail
et al.\ 2000a) and/or mm interferometry (Downes et al.\ 1999; Bertoldi
et al.\ 2000) as well as near-infrared observations are required in
order to uncover the proper counter-part.  The galaxy SMM\,J00266+1708
is an excellent example of such a source.

\section{OBSERVATIONS}

\subsection{Background Data on SMM\,J00266+1708}

The sub-mm galaxy SMM\,J00266+1708 is the second brightest galaxy in the
SCUBA Cluster Lens Survey (Smail et al.\ 1998).  Initially, the source
was tentatively associated with a possible interacting pair of galaxies,
M1 and M2, revealed by deep $I$-band imaging ($3\sigma=26.1$\,mag) with
the {\it Hubble Space Telescope} [{\it HST}] (Smail et al.\ 1998).
Spectroscopy of the galaxy M2 showed a bright [O{\sc ii}] emission line
at $z=1.226$, consistent with a luminous star-forming galaxy (Barger et
al. 1999).  In the fall of 1998, we searched for redshifted
CO(2$\rightarrow$1) emission corresponding to the redshift of M2 at the
Owens Valley Millimeter Array\footnote{The Owens Valley Millimeter Array
is a radio telescope facility operated by the California Institute of
Technology and is supported by NSF grant AST 9981546.} (OVRO).  We
failed to detect any CO emission at the redshift of M2; $S({\rm CO}) <
1.3$\,Jy$\kps$ ($3\sigma$), assuming a standard $300\kps$ line width.
If M2 was the correct counter-part and the source was similar to the
sub-mm galaxies with previous CO detections (Frayer et al. 1998, 1999),
we would have expected a CO(2$\rightarrow$1) line strength of
approximately $7\,{\rm Jy}\kps$.  The nondetection of CO questions the
association of M2 with the sub-mm galaxy.  Additional optical
spectroscopy showed that M1 is at the redshift of the foreground cluster
($z=0.39$), and hence, M1 and M2 are not an interacting pair after all
(Barger et al. 1999).  These results further bring into doubt the
initial association of M1 and M2 with the sub-mm emission based on
optical morphology alone.

Besides M1 and M2, the only other optically visible source spatially
consistent with the SCUBA position is M8 (Fig. 1).  However, the galaxy
M8 shows no unusual properties that would suggest an association with a
luminous sub-mm source.  More significantly, the field has a weak radio
source (Smail et al.\ 2000a) whose position is consistent with the SCUBA
source, but is slightly offset from M8.  Since the brightest sub-mm
galaxies tend to have radio counter-parts (Smail et al.\ 2000a; Barger,
Cowie, \& Richards 2000), we could reasonably expect an association
between the sub-mm galaxy and the radio source.  If this is the case,
the radio emission of SMM\,J00266+1708 is too weak to be consistent with
a redshift of $z=0.44$ for M8 (Barger et al.\ 1999), based on the
redshift dependency predicted for the sub-mm/radio spectral index
(Carilli \& Yun 1999).  Although the sub-mm/radio spectral index only
provides an estimate of the redshift given the uncertainties of source
temperature and properties (Blain 1999), it does provide a powerful
technique for discriminating between low-redshift ($z\la0.5$) and
high-redshift ($z\ga2$) galaxies.  SMM\,J00266+1708 is expected to be at
redshifts $z>2$ since its sub-mm/radio spectral index of $\alpha=1$ is
much larger than any known galaxy at low redshift (Smail et al.\ 2000a).
Therefore, it is unlikely that M8 is the sub-mm counter-part.  Since
none of the optically detected galaxies are plausibly associated with
the sub-mm emission, we have carried out mm-continuum and near-infrared
observations of SMM\,J00266+1708.

\subsection{OVRO 1.3\,mm Continuum Observations}

We have taken sensitive 1.3\,mm interferometric observations of
SMM\,J00266+1708 in order to accurately constrain the position of the
sub-mm source.  SMM\,J00266+1708 was observed several times using the
OVRO array in 1999.  Approximately 10 hours of on source data were
obtained during good conditions in the low resolution configurations of
the array (baseline lengths ranging from 15m to 119m).  We observed with
four separate 1~GHz continuum bands centered at 229.0, 230.5, 233.5, and
235.0 GHz.  The 4~GHz of total continuum bandwidth represents a factor
of two increase in bandwidth over what was previously achievable at
OVRO.

%
\includegraphics{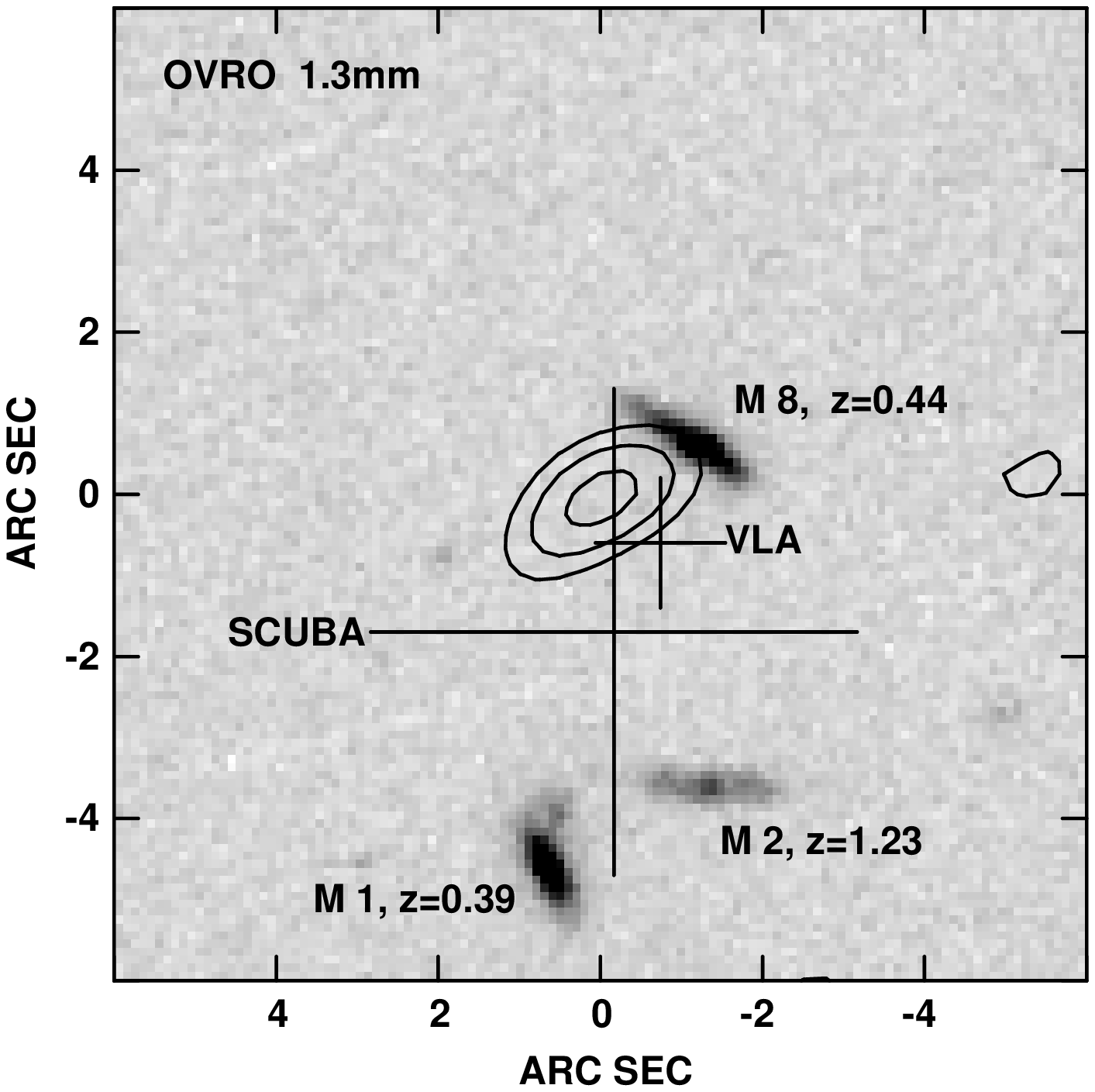}
\vspace*{3.8in}

{\footnotesize {\sc Fig}.~~1.--- The OVRO 1.3\,mm contour map overlaid on
the {\it HST} $I$-band optical image (Smail et al.\ 1998). The rms level is
1.1\,mJy/beam, and the contour levels are $1\sigma\times (-3,3,4,5)$.
The source is unresolved by the $2\arcsec$ OVRO beam.  The position of
the 21\,cm radio source (Smail et al.\ 2000a) is shown by the cross
labeled ``VLA'', while the positional uncertainty of the sub-mm source
is marked by the cross labeled ``SCUBA''.}
\vspace*{4mm}

We observed the bright quasar 3C454.3 every 20 minutes for amplitude and
phase calibration.  Since 3C454.3 is 22 degrees from SMM\, J00266+1708,
we also interweaved observations of the nearby quasar 0007+106 (B1950.0)
in order to test the quality of the calibration.  The systematic
positional uncertainty of the data is estimated to be better than
$0\farcs3$.  Observations of the planets Uranus and Neptune were used to
calibrate the absolute flux scale.  By using the flux history of
3C454.3, we estimate a flux calibration uncertainty of 20\% for the
data.

We combined the data for the four individual 1~GHz bands to produce the
1.3\,mm detection at a mean frequency of 232.0 GHz.  The contours in
Figure~1 show the resultant natural--weighted image. No primary beam
correction was required since the source was located within $4\arcsec$
(1/8 of the primary beam width) of the phase center.  We made no
correction for possible variations of source strength across the four
individual bands.  These variations are expected to be less than 10\%,
assuming dust emission, and were undetected within the uncertainties of
the data.

\subsection{Keck $K$-band Imaging}

After the 1.3\,mm detection, we obtained $K$-band ($2.2\mu$m) data to
search for the galaxy responsible for the sub-mm emission at the
position derived from the OVRO data.  We observed SMM\,J00266+1708 using
the Near Infrared Camera (NIRC) on Keck~I\footnote{The W.\ M.\ Keck
Observatory is operated as a scientific partnership among the California
Institute of Technology, the University of California, and the National
Aeronautics and Space Administration.  The Keck Observatory was made
possible by the generous financial support of the W.\ M.\ Keck
Foundation.} on UT 1999 October 01.  NIRC is a $256\times256$ pixel InSb
detector with a pixel scale of $0\farcs15$ (Matthews \& Soifer 1994).
We observed using the standard $K$-band filter instead of the bluer
$K_{\rm s}$ filter since the object is expected to be red.  Integrations
were taken using $10\times 6$ second coadds, and we randomly dithered
the integrations within a $8\arcsec\times8\arcsec$ box to provide
uniformity across the image.  We obtained a total of 4.3 hours of data
on source.  The seeing-disks of the stars observed throughout the night
varied from $0\farcs3$ to $1\arcsec$ (FWHM).

In reducing the data, a combined set of dark frames was subtracted from
each individual exposure to remove the dark current as well as the bias
level.  The dark-subtracted exposures were divided by a normalized
skyflat which was generated from the on object exposures themselves.
Frames were sky-subtracted using the temporally--adjacent images to
produce the reduced exposures.  The individual reduced exposures were
aligned to the nearest pixel using common objects in the frames.  We
observed a set of near-infrared standard stars (Persson et al.\ 1998) at
a range of airmasses in order to correct the data for extinction.  The
uncertainty of the derived magnitude scale is estimated to be better
than 0.04 magnitude, based on the dispersion in the zero-points derived
throughout the night.

{\small

\begin{deluxetable}{lcccl}
\tablenum{1}
\tablecaption{Observed Properties of SMM\,J00266+1708}
\tablewidth{350pt}
\tablehead{\colhead{Wavelength}&\colhead{$\alpha$(J2000)}
&\colhead{$\delta$(J2000)}&\colhead{Flux\tablenotemark{a}}
&\colhead{Telescope}\nl              
\colhead{ }&\colhead{$00^{\rm h}26^{\rm m}$}&
\colhead{$+17\arcdeg 08\arcmin$}& \colhead{ }&\colhead{ }}
\startdata

21.4\,cm & $34\fs06\pm0\fs06$ & $33\farcs1\pm0\farcs8$ &
$94\pm15 \muJy$ & VLA \nl

3.54\,cm & ... & ... & $<34\muJy (3\sigma)$ & VLA \nl 

1.29\,mm & $34\fs10\pm0\fs04$ & $33\farcs7\pm0\farcs5$ &
$6.0\pm1.7$\,mJy & OVRO$^{\rm b}$ \nl

850\,$\mu$m & $34\fs1\pm0\fs2$ & $32\arcsec\pm3\arcsec$ &
$18.6\pm2.4$\,mJy & JCMT \nl

450\,$\mu$m & ... & ... &
$<60$\,mJy $(3\sigma)$ & JCMT \nl 

2.2\,$\mu$m&  $34\fs11\pm0\fs07$ & $33\farcs2\pm1\farcs0$ & 
$0.68\pm0.07\muJy$ & Keck$^{\rm b}$ \nl

& & & $(K=22.45\pm0.11)$& \nl

0.8\,$\mu$m& ... & ... & 
$<0.09\muJy$ ($3\sigma$) & HST \nl

& & & $(I<26.1)$ & \nl

\enddata 
\tablenotetext{a}{Flux uncertainties include both systematic
and rms errors.\\ \hspace*{3.mm}$^{\rm b}$Observations presented in
this paper.}
\vspace*{-4mm}
\end{deluxetable}
}

\section{RESULTS}

\subsection{Images}

Figure~1 shows the 1.3\,mm continuum map.  The source was detected at
the $5\sigma$ level and was unresolved with a synthesized beam size of
$\theta_{b}=2\farcs3 \times 1\farcs9$.  At the observed frequency of 232
GHz (1.29\,mm), we derive a flux density of $6.0\pm1.1$\,mJy by fitting
a Gaussian to the peak of the emission.  The total uncertainty in the
flux is 28\% which includes the 20\% calibration error and the rms error
of 20\% combined in quadrature.  The strength of the 1.3\,mm source is
consistent with extrapolating the thermal dust spectrum expected from
the sub-mm source (Fig.~3).  We conclude that the 1.3\,mm source is
associated with the SCUBA source and is likely representative of the
bulk of the sub-mm emission.

Figure~2 shows the $K$-band image.  We detected a new galaxy $1\farcs5$
south-east of the nucleus of M8 at the position of the OVRO 1.3\,mm
continuum source.  Since the seeing varied significantly throughout the
night, we combined only the best 2.4 hours of data to produce the
$K$-band image.  After convolving the data with a 2 pixel (FWHM)
Gaussian to reduce pixel-to-pixel variations, the resolution of data is
$0\farcs5$ (FWHM), including seeing.  The image has a $1\sigma$ limiting
$K$-band surface brightness of $\mu =24.8\,{\rm mag}/\Box\arcsec$ or
$0.04\mu$Jy/beam, adopting 646\,Jy for the flux density equivalent for a
zero $K-$band magnitude (Neugebauer et al.\ 1987).

%
\includegraphics{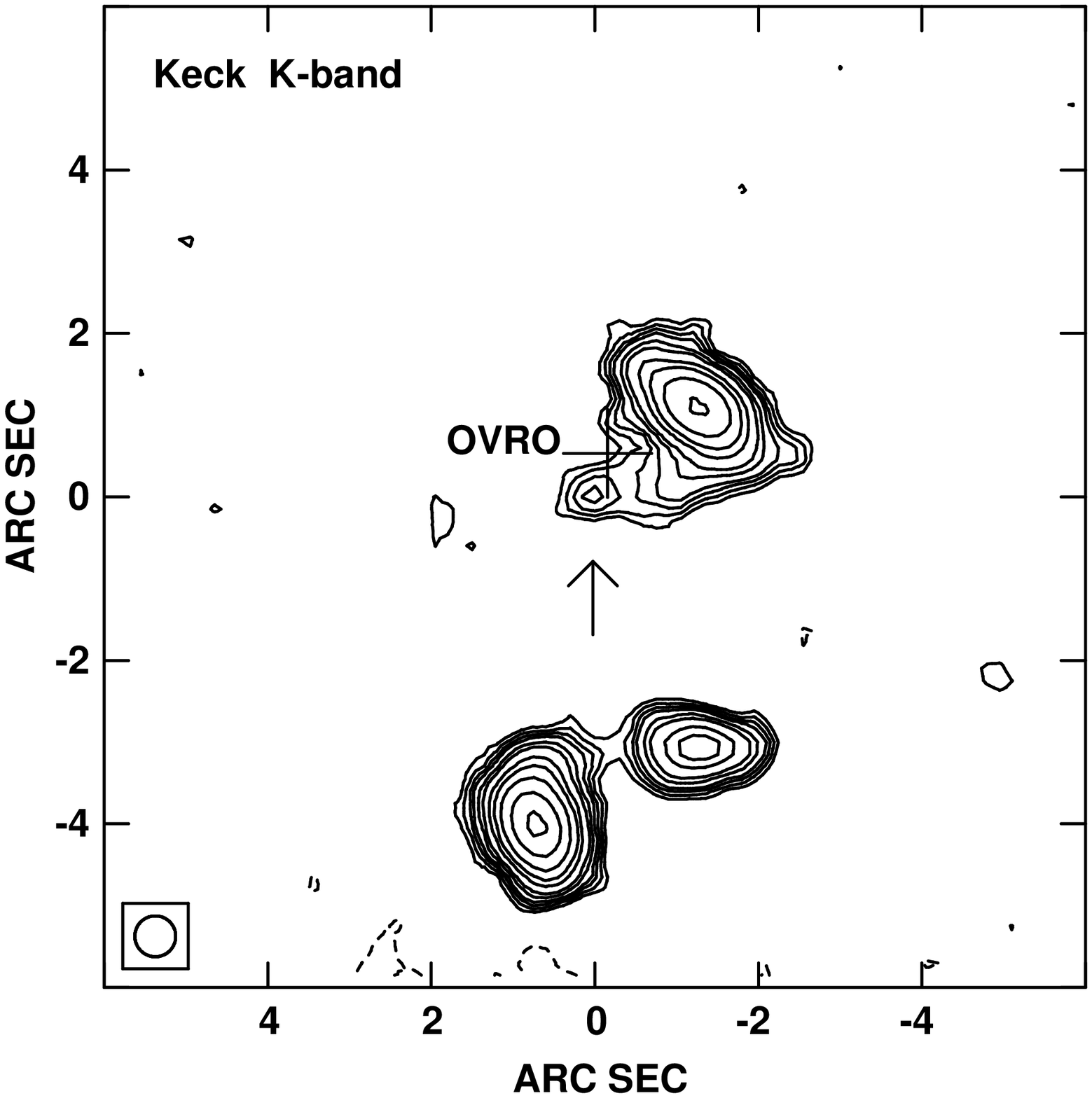}
\vspace*{3.9in} {\footnotesize {\sc Fig}.~~2.--- The Keck $K$-band ($2.2\mu$m)
image.  The arrow points to the new galaxy thought to be the
counter-part of SMM\,J00266+1708.  The rms of the image is 24.8
mag/$\Box\arcsec$ ($0.04\mu$Jy/beam), and the contours are
$1\sigma\times (-3,3,4,5,6,8,10,15,20,30,50,80)$.  The seeing disk
(beam) of the NIR data is shown in the lower left
($0\farcs5\times0\farcs5$).  The position of the 1.3\,mm source is shown
by the cross labeled ``OVRO''.}\vspace*{4mm}

Photometry on the galaxy was done using an aperture to sum up the
emission from the regions around the south-eastern peak that are well
separated from M8.  The integrated emission for the galaxy is detected
at about the $10\sigma$ level with a magnitude of $K=22.45\pm0.11$
($0.68\pm 0.07\muJy$, $K_{\rm AB}\simeq 24.3$).  There is additional
emission westward from the south-eastern peak and near M8.  There is no
evidence in the $I-$band image for this extension.  If we assume that
this emission is associated with the south-eastern peak, we would derive
a total magnitude of $K=21.5$ after subtracting the bright disk of M8
from the image.  For the remainder of the paper, we neglect this
additional emission since it is unclear what fraction is associated with
M8, the south-eastern peak, or neither.

\subsection{Astrometry}

The $K$-band image was registered to the {\it HST} optical image on the
APM coordinate system using the galaxies M1 and M8.  The optical
astrometry in the {\it HST} image has an rms uncertainty of only
$0\farcs2$.  However, the dominant source of registration error in
comparing the $K$-band data with the radio and 1.3\,mm data is due to
the systematic offsets between the APM system and the radio reference
frame (Johnston et al. 1995).  These offsets may be as large as
$1\arcsec$ which we adopt as a conservative estimate of the total error
in the $K$-band position.  The positional error for the 1.3\,mm source
is $0\farcs5$.  This includes a systematic astrometry uncertainty of
$0\farcs3$ from calibration combined in quadrature with the statistical
error related to the signal--to--noise (S/N) of approximately $0\farcs4$
($\theta_{b}/[S/N]$).  The 21\,cm radio data have a synthesize beam size
of $5\arcsec$ and a positional error of about $0\farcs8$
($\theta_{b}/[S/N]$).  The 850$\mu$m source detected by SCUBA has a
positional accuracy of approximately $3\arcsec$ (Smail et al. 1998).

The positions of SMM\,J00266+1708 measured at 21\,cm, 1.3\,mm,
850$\mu$m, and $2.2\mu$m are all consistent with each other within the
errors (Table~1).  The positional coincidence alone suggests an
association between the sources at different wavelengths.  The
probability of randomly finding a 22.5 mag $K$-band galaxy within a 1
square-arcsec box is only 1\%, based on the observed $K$-band surface
densities of galaxies (Djorgovski et al.\ 1995; Moustakas et al.\ 1997).
By taking into account the red nature of the galaxy ($I-K>3.6$), the
likelihood of a random association decreases even more.  Only about 10\%
of faint galaxies ($22.0 \leq K \leq 22.9$) have $(I-K)>3.5$ (Moustakas
et al.\ 1997).  Hence, the probability of a chance association of such a
faint red galaxy with the 1.3\,mm source is only $10^{-3}$.  An even
stronger case can be made on the association between the radio source
and the 1.3\,mm source.  Radio source counts (Richards et al.\ 1999;
Richards 2000) imply the probability of randomly finding such a strong
radio source within 1 square arcsec of the 1.3\,mm source is only about
$10^{-5}$.  We conclude the sources detected at 21\,cm, 1.3\,mm,
850$\mu$m, and $2.2\mu$m are all likely associated with each other.

\subsection{Spectral Energy Distribution and Redshift}

Table~1 presents the flux density measurements and upper-limits observed
for SMM\,J00266+1708.  The 21\,cm radio data are presented by Smail et
al.\ (2000a).  We measure a flux density of $94\pm15 \muJy$ for the
21\,cm radio source by fitting a Gaussian to the unresolved emission.
We also report a 3.5\,cm upper-limit for the galaxy based on sensitive
observations of the cluster (A. R. Cooray, private communication).  The
sub-mm flux at 850$\mu$m has been previously tabulated by Barger et al.\
(1999), and the 450$\mu$m upper limit is discussed by Smail et al.\
(2000b).  The optical $I$-band limit of 26.1\,mag is derived from the
observations presented by Smail et al.\ (1998).  The 1.3\,mm and
$K$-band measurements are based on the observations presented in this
paper.

As stated earlier (\S 2.1), SMM\,J00266+1708 is thought to lie at a high
redshift of $z\ga2$ based on the sub-mm/radio spectral index of the
galaxy (Carilli \& Yun 1999; Smail et al.\ 2000a).  High redshifts are
also required to account for the low 450$\mu$m/850$\mu$m ratio, assuming
dust emission with properties similar to that found in low-redshift
luminous starbursts (Dunne et al.\ 2000).  The optical and NIR imaging
data by themselves provide little constraint on the redshift of
SMM\,J00266+1708 given the wide range of optical/NIR properties found
for the sub-mm population (e.g., Ivison et al.\ 2000a).  The best
redshift constraints for SMM\,J00266+1708 come by fitting the spectral
energy distribution (SED) of the galaxy from radio to sub-mm
wavelengths.  Figure~3 shows the observed SEDs of the low redshift ULIGs
Arp\,220 and Mrk\,231 (Rigopoulou, Lawrence, \& Rowan-Robinson 1996), as
well as the $z=1.44$ extremely red galaxy (ERO) HR10 (Dey et al.\ 1999),
the relatively blue sub-mm starburst SMM\,J14011+0252 (Ivison et al.\
2000a), and the sub-mm galaxy SMM\,J02399-0136 which contains an AGN
(Ivison et al.\ 1998).  These ULIG systems are chosen for comparison
since they represent the reasonable range of possible SEDs expected for
the sub-mm population.  Assuming a likely range of dust temperatures
(30--50\,K) and fitting all the data to the nearest 0.5 redshift unit,
we estimate a redshift of $z=3.5\pm1.5$ for SMM\,J00266+1708.

The wide range of optical flux densities and colors for ULIGs/sub-mm
galaxies (Fig. 3) demonstrates the difficulty of attempting to estimate
the far-infrared luminosities of ultraluminous galaxies based solely on
optical observations (c.f., Adelberger \& Steidel 2000).  Bolometric
luminosities estimated from sub-mm flux densities are significantly more
robust since the observed emission from high-redshift galaxies arises
near the peak of the SED.

%
\includegraphics{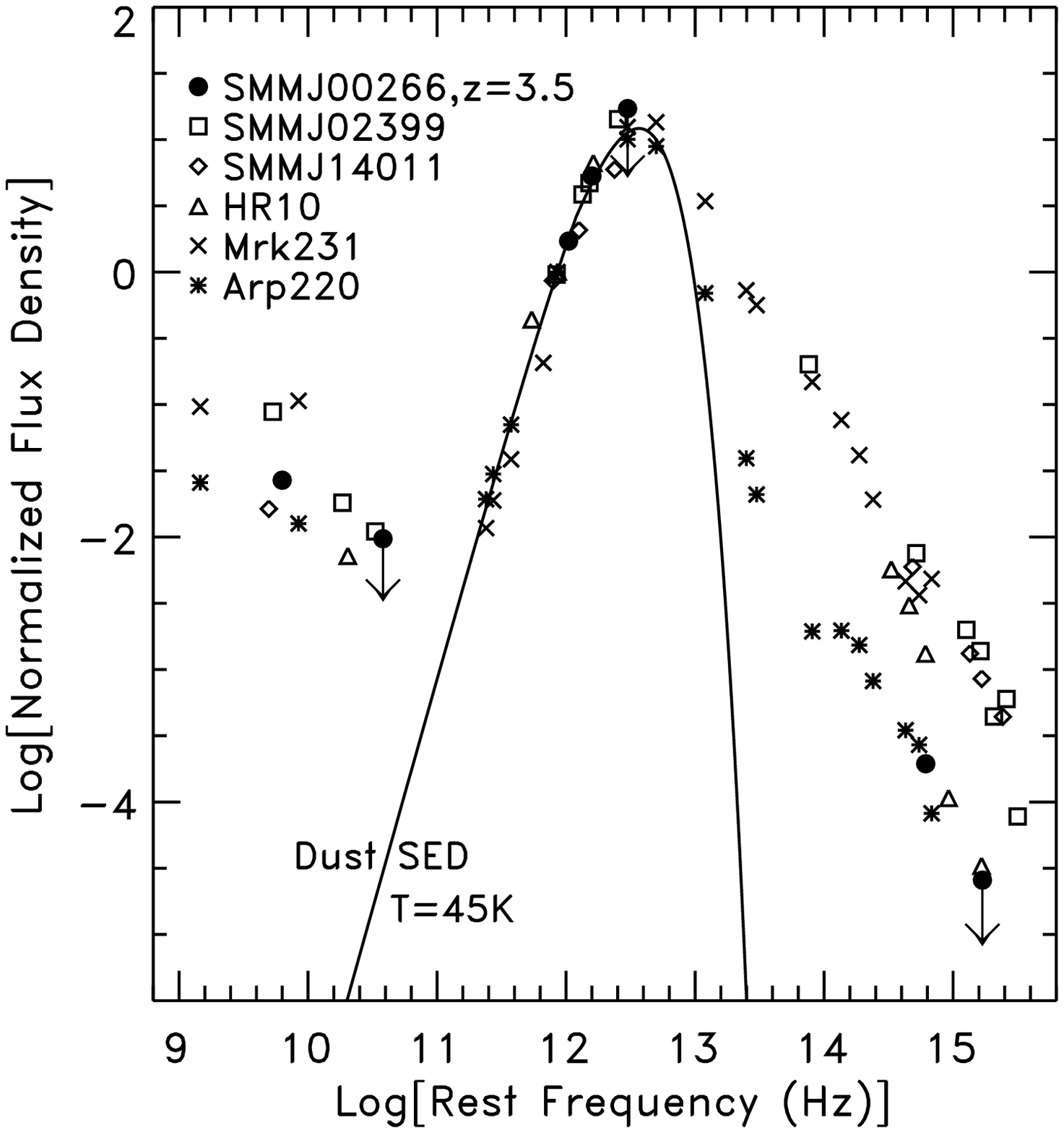}
\vspace*{4.0in} {\footnotesize {\sc Fig}.~~3.--- The SED of
SMM\,J00266+1708 (solid circles) adopting the best fit redshift of
$z=3.5\pm1.5$ compared to other ULIG systems.  All data have been
normalized to a rest wavelength of 350$\mu$m.  The upper-limits for
SMM\,J00266+1708 are marked with downward arrows.  The solid line shows
an example SED for dust emission assuming a typical temperature of
$T=45$\,K and $\beta=1.5$ for the power-law dependence of the dust
absorption coefficient.}

\subsection{Lensing}

In this section, we estimate the effect of gravitational lensing.
SMM\,J00266+1708 is amplified by the potential of the $z=0.39$
foreground cluster Cl\,0024+16.  Assuming a redshift of $z=1.23$, Barger
et al.\ (1999) find an amplification factor of 1.6 due to the cluster.
Adopting a more realistic source redshift of $z\sim3.5$, the
amplification factor is increased to 2.1 using the relationships given
by Schneider, Ehlers, \& Falco (1992) to rescale the appropriate angular
size distances.  The amplification factor due to the cluster is fairly
insensitive to redshift for $z\ga2$.

Since SMM\,J00266+1708 is near the line of sight of the galaxy M8
($z=0.44$), we could expect additional gravitational lensing due to M8.
To estimate this effect, we approximate both M8 and SMM\,J00266+1708 as
point sources and use the equations for a simple Schwarzschild lens
(Schneider et al.\ 1992).  M8 appears to be a normal spiral galaxy with
a typical absolute $V$-band magnitude of $M_{V}=-20$ (Smail et al.\
1997).  Assuming a total mass of $2\times10^{11} M_{\sun}$ within the
central 10\,kpc of M8 (which corresponds to the angular separation of
$1\farcs5$ between SMM\,J00266+1708 and M8), we estimate that
SMM\,J00266+1708 is amplified by a factor of 1.14 due to M8.  Including
lensing by both the cluster and M8, we expect a total magnification
factor of about 2.4 ($\pm0.5$) for SMM\,J00266+1708.

SMM\,J00266+1708 is the second sub-mm galaxy (ERO-N4 is the other
example [Smail et al.\ 1999]) found in the Cluster Lens Survey thought
to be lensed by a low redshift spiral galaxy.  These results may suggest
that lensing due to galaxies may play a significant role for the
brightest sub-mm galaxies.  The current observations of sub-mm counts
indicate that future mm/sub-mm surveys covering large areas of the sky
should detect many strongly lensed sources (Blain, M\"{o}ller, \& Maller
1999).

\section{DISCUSSION}

We have concentrated our follow-up studies on the nine brightest
galaxies detected in the sub-mm Cluster Lens Survey (Smail et al.\
1998).  Currently, only three of the nine galaxies have optical
counter-parts with redshifts.  Two of these have been confirmed by CO
observations at OVRO (Frayer et al.\ 1998, 1999), and the third is a
Seyfert ring galaxy at $z=1.06$ recently detected in CO with the IRAM
interferometer (Soucail et al.\ 1999; Kneib et al.\ 2000).

At least four of the nine systems are undetected at optical wavelengths
($I>26$--27 after correcting for lens amplification) and have only been
detected with $K$-band imaging.  Two of these are bright enough
($K=19.1$, 19.6\,mag) to be classified as EROs (Smail et at.\ 1999).  An
additional faint ($K=21$, $R>26$) galaxy was found associated with a
relatively bright (500 $\mu$Jy) radio counter-part (Ivison et al.\
2000a, 2001).  SMM\,J00266+1708 is the faintest sub-mm counter-part
found to date at $K=22.5$.  Only two of the nine sources still require
deep $K$-band imaging and have uncertain counter-parts.  Depending on
the results for these last two unknown systems, the data suggest that
approximately 40\%--70\% (4/9--6/9) of the sub-mm population as a whole
are very faint/red galaxies which are undetected at optical wavelengths.
It is still unclear what fraction of the sub-mm galaxies are EROs
[$R-K>6$] (Thompson et al.\ 1999).  At least two out of nine ($\ga20$\%)
galaxies in the Cluster Lens Survey are EROs (Smail et al.\ 1999), and
other sub-mm surveys have uncovered additional faint EROs associated
with SCUBA sources (Ivison et al.\ 2000b).  Future, much deeper optical
observations may show that a high fraction of sub-mm galaxies galaxies
are EROs.

\subsection{Intrinsic Properties of SMM\,J00266+1708}

SMM\,J00266+1708 is an extremely faint galaxy.  After correcting for
lensing, the source-frame magnitude of the galaxy is $K=23.4$.  We adopt
a lensing magnification of 2.4, a redshift of $z=3.5$, and a cosmology
of $H_o=50\kps$\,Mpc$^{-1}$ and $q_o=0.5$ in order to estimate the
properties of SMM\,J00266+1708.  Given the redshift uncertainty, we
cannot accurately determine many of the intrinsic properties of
SMM\,J00266+1708.  One exception is the bolometric sub-mm/far-infrared
(FIR) luminosity which is relatively insensitive to redshift for $z\ga1$
(Blain \& Longair 1993).  From the 850$\mu$m measurement, the implied
intrinsic FIR luminosity for SMM\,J00266+1708 is $L({\rm FIR}) \simeq
10^{13} L_{\sun}$.  This corresponds to a star-formation rate of massive
stars ($M>5M_{\sun}$) of approximately 500--900 $M_{\sun}$\,yr$^{-1}$
(Scoville et al.\ 1997; Condon 1992).  These results are consistent with
estimates for other sub-mm galaxies (Ivison et al.\ 1998, 2000a).  It is
still not known whether or not an AGN contributes significantly to the
bolometric luminosity of SMM\,J00266+1708.  If an AGN is present, it is
not a strong radio source and must be heavily obscured at optical
wavelengths (Fig.~3).  Based on its SED, SMM\,J00266+1708 appears most
similar to highly-reddened ULIG/starbursts such as Arp~220 or HR10 at a
redshift of $z\sim 3.5$.

\subsection{Comparison of Sub-mm and Lyman-break Populations}

The relative importance of the sub-mm and Lyman-break galaxies to the
global star-formation rate at high-redshift is an active area of
discussion (Hughes et al.\ 1998; Guiderdoni et al.\ 1998; Blain et
al. 1999b, 1999c; Trentham et al. 1999; Peacock et al. 2000).  Depending
on the exact contribution of AGN in the sub-mm population, the
ultraluminous sub-mm galaxies ($>10^{12} L_{\sun}$;
$S[850\mu$m$]\ga1$\,mJy) account for approximately 30--50\% of the total
amount of star-formation at high-redshift (Blain et al.\ 1999b).  It is
truly remarkable that such a high fraction of all star-formation at
high-redshift occurs in ULIG/sub-mm galaxies.  In contrast, ULIGs only
contribute about 0.2\% of the total amount of star formation in the
local universe, based on the survey of Kim \& Sanders (1998).  Hence,
the evolution in the amount of star formation occurring in ULIGs is
about 100 times stronger than the global increase of the star-formation
rate at high redshift seen for all galaxies (Madau et al.\ 1996; Steidel
et al.\ 1999).

It has been suggested for both the sub-mm and Lyman-break populations
that each represents the formative phases of massive $L^{*}$ galaxies
(Blain et al.\ 1999c; Giavalisco et al.\ 1998).  In this scenario, the
sub-mm systems may be associated with a very luminous, short-lived and
heavily dust enshrouded starburst, while the Lyman-break galaxies would
be associated with a longer-lived, less luminous phase of star formation
(Blain et al.\ 1999c).  If this scenario is correct, we could expect
massive reservoirs of molecular gas associated with both populations.
The detection of massive reservoirs of molecular gas (Frayer et al.\
1998, 1999) suggests that the sub-mm population is associated with
gas-rich massive galaxies ($\ga L^{*}$).  The molecular gas masses of
the sub-mm galaxies are 10--50 times greater than that found for the
Milky Way.  In contrast, the best studied Lyman-break galaxy cB58 (Yee
et al.\ 1996; Pettini et al.\ 2000; Teplitz et al.\ 2000) has even less
molecular gas than the Milky Way, after correcting the observed CO
upper-limit for lensing and suspected metallicity effects (Frayer et
al.\ 1997).  Therefore, the Lyman-break sources may be sub-$L^{*}$
systems representing the building blocks of more massive galaxies, as
suggested by Lowenthal et al.\ (1997).

Most sub-mm galaxies are similar to SMM\,J00266+1708 in being too faint
and/or too red to be included in the Lyman-break surveys (e.g., Steidel
et al.\ 1999).  The one notable exception is the sub-mm selected galaxy
pair, SMM\,J14011+0252 J1/J2 (Ivison et al.\ 2000a; Adelberger \& Steidel
2000), but for this system most of the blue light arises from J2, while
the bulk of the bolometric luminosity is thought to be due to the much
redder J1 component (Ivison et al.\ 2001).  Due to the high levels of
dust obscuration for the sub-mm galaxies, there is very little overlap
between the sub-mm selected galaxies and optically-selected Lyman-break
systems.  Even for the Lyman-break galaxies, most of their
star-formation ($\sim$80\%) is obscured by dust at observed optical
wavelengths (Peacock et al.\ 2000; Adelberger \& Steidel 2000).  Hence,
it is clear that most of the star-formation activity at high-redshift is
hidden from view at optical/ultraviolet wavelengths.

The results presented for SMM\,J00266+1708 have important implications
on our general understanding of star-formation at high redshift.
Roughly half of the total amount of star formation at high redshift is
thought to occur in the sub-mm galaxies (Blain et al.\ 1999b), while the
other half is inferred from the optically--selected Lyman-break sources
(Peacock et al.\ 2000).  Since about 50\% of the sub-mm galaxies are
undetected at optical wavelengths, a significant fraction ($\sim 25$\%)
of the total amount of high-redshift star formation may occur in the
very faint/red sub-mm galaxies similar to SMM\,J00266+1708.  Similar
conclusions have also been reached for a radio-selected sample of sub-mm
galaxies with faint optical counter-parts (Barger et al.\ 2000).  The
fact that many of the most luminous high-redshift starbursts/AGN are too
faint to be studied at optical wavelengths highlights the importance
that future sensitive mm--NIR wavelength instruments will have on our
understanding of the distant universe.

\section{CONCLUDING REMARKS}

We report the identification of the sub-mm source SMM\,J00266+1708 with
a faint red galaxy ($K=22.5$\,mag) which is undetected at optical
wavelengths, despite very deep observations.  This source has an
extremely high luminosity of approximately $10^{13}L_{\sun}$ even after
correcting for lensing.  The current data for the sub-mm Cluster Lens
Survey suggest that 40\%--70\% of the sub-mm population as a whole are
faint/red galaxies which are undetected at optical wavelengths.  These
faint/red sub-mm galaxies are thought to contribute significantly to the
total amount of star formation at high redshift and are hence important
to our understanding of the early evolution of galaxies.

The redshift of SMM\,J00266+1708 is currently unknown, but the galaxy is
expected to be at a redshift $z>2$.  Obtaining a redshift will be
extremely challenging with current instrumentation.  At $K=22.5$ mag,
the galaxy pushes the capabilities of even the largest ground based
telescopes.  Since the galaxy is relatively red ($I-K > 3.6$), we expect
H$\alpha$ to be the brightest optical emission line, and perhaps the
only optical line currently detectable based on comparisons with the ERO
HR10 (Dey et al.\ 1999).  If the redshift is similar to that estimated
from the SED of the galaxy ($z\sim3.5$), H$\alpha$ would be shifted
redward of $K$-band, making ground based observations extremely
challenging.

We could expect much fainter $K-$band magnitudes of 23--26 for similar
ULIG/sub-mm galaxies ($\ga 10^{12} L_{\sun}$) which are unlensed.
Therefore, obtaining optical/NIR redshifts for many of the sub-mm
galaxies may have to wait for the {\it Next Generation Space Telescope}.
Alternatively, redshifts could be directly measured from the CO lines
themselves with future ground-based instruments, such as the Atacama
Large Millimeter Array (ALMA), operating at mm-wavelengths (Blain et
al.\ 2000).  Sensitive interferometric observations at
sub-mm/mm-wavelengths with the next generation of instruments will be
crucial for our understanding of these dust-obscured systems.

\acknowledgements

We thank the staff at the Owens Valley Millimeter Array and the Keck
Observatory who have made these observations possible.  We thank Andrew
Blain and Jean-Paul Kneib for useful discussion and their work on the
SCUBA Lens Survey Sample.  We thank Frazer Owen, Glenn Morrison, and
Asantha Cooray for their work on the radio data.  We thank Dave Thompson
for his set of IRAF tasks called NIRCtools which were used to help
reduce the NIR data.

\end{document}